\documentclass[twocolumn]{aastex63}

\usepackage{graphicx}
\usepackage{amsmath}

\let\tablenum\relax
\usepackage{siunitx}

\usepackage{xcolor}
\usepackage[percent]{overpic}
\usepackage{ulem}

\usepackage[shortcuts, acronym]{glossaries-extra}
\usepackage{glossaries-prefix}
\setabbreviationstyle[acronym]{long-short}
\glsdisablehyper
\newacronym{PIC}{PIC}{particle-in-cell}
\newacronym{NTPA}{NTPA}{nonthermal particle acceleration}
\newacronym[prefixfirst={a\ }, prefix={an\ }]{FP}{FP}{Fokker-Planck}
\newacronym{MHD}{MHD}{magnetohydrodynamic}
\newacronym{AGN}{AGN}{active galactic nuclei}
\newacronym{PWN}{PWN}{pulsar wind nebulae}
\newacronym{GRB}{GRB}{gamma ray bursts}
\newacronym{3D}{3D}{three-dimensional}

\newcommand{\fx}{f(\gamma)}
\newcommand{\fxt}{f(\gamma, t)}

\newcommand{\pdx}{\partial_\gamma}
\newcommand{\pdt}{\partial_t}

\newcommand{\ExB}{\vec{E \times B}}
\newcommand{\EsqBsq}{E^2 +\nobreak B^2}

\newcommand{\pref}{D}
\newcommand{\vPrefMag}{v_\pref}
\newcommand{\vPref}{\vec{v}_\pref}

\newcommand{\nba}[1]{#1\nobreak}
\newcommand{\nbaeq}{\nba{=}}
\newcommand{\nbasim}{\nba{\sim}}
\newcommand{\nbasimeq}{\nba{\simeq}}
\newcommand{\nbaequiv}{\nba{\equiv}}
\newcommand{\nbaapprox}{\nba{\approx}}

\newcommand{\Dcoeff}[1][]{D#1}
\newcommand{\Acoeff}[1][]{A#1}
\newcommand{\Mcoeff}[1][]{M#1}
\newcommand{\Dcontrib}{\pdx \Dcoeff}
\newcommand{\Aeqn}{\Acoeff(\gamma) \equiv \Mcoeff - \partial_\gamma \Dcoeff}

\newcommand{\bcGamma}{{\gamma_0}}

\newcommand{\mean}{\overline{\gamma}}
\newcommand{\stdev}{\delta\gamma_\rms}
\newcommand{\relStdev}{\stdev / \bcGamma}
\newcommand{\relMean}{\mean / \bcGamma}

\newcommand{\changeT}{\Delta t}
\newcommand{\sqrtT}{\sqrt{\changeT}}
\newcommand{\stdevProptoSqrt}{\stdev(\changeT) \propto \nobreak \sqrtT}

\newcommand{\gammaPeak}{\gamma_\textrm{peak}}
\newcommand{\gammaAvg}{\gamma_\textrm{avg}}

\newcommand{\tLarmor}{T_{\rm L}}
\newcommand{\tLarDef}[2]{2 \pi {#1} \me c / e {#2}}

\newcommand{\figref}[1]{Fig.~\ref{#1}}
\newcommand{\figsref}[2]{Figs.~\ref{#1} and~\ref{#2}}
\newcommand{\eqnref}[1]{Eq.~\eqref{#1}}
\newcommand{\eqnsref}[2]{Eqs.~\eqref{#1} and~\eqref{#2}}

\newcommand{\invSqrt}[1]{(#1)^{-1/2}}
\newcommand{\invLC}{c/L}

\newcommand{\teq}[1]{t\nbaeq{}#1}

\newcommand{\rms}{\textrm{rms}}
\newcommand{\init}{\textrm{init}}

\newcommand{\me}{m_e}
\newcommand{\restEnergy}{\me c^2}
\newcommand{\relEnergy}[1]{\gamma{#1} \restEnergy}

\newcommand{\mfp}{\lambda_\textrm{mfp}}
\newcommand{\Do}{D/\gamma^2}%{D_0}%

\newcommand{\gyro}[1]{gyro-{#1}}

\newcommand{\MaxwellJuttner}{Maxwell-J\"{u}ttner}
\newcommand{\Alfven}[1]{Alfv\'en{#1}}

\newcommand{\nSimPtcls}{2.4e11}
\newcommand{\nTrackPtcls}{8e5}
\newcommand{\startTL}{10.0 L/c}
\newcommand{\Dmeas}{0.06}
\newcommand{\lowPower}{2/3}

\renewcommand{\vec}[1]{\mathbf{#1}}
\newcommand\unitvec[1]{\vec{\hat{#1}}}

\renewcommand\cite\citep

\begin{document}
\title{First-principles demonstration of diffusive-advective particle acceleration in kinetic simulations of relativistic plasma turbulence}

\author{Kai Wong}
\affiliation{Center for Integrated Plasma Studies, Physics Department, 390 UCB, University of Colorado, Boulder, CO 80309, USA}

\author{Vladimir Zhdankin}
\altaffiliation{NASA Einstein fellow}
\affiliation{Department of Astrophysical Sciences, Princeton University, Peyton Hall, Princeton, NJ 08544, USA}

\author{Dmitri A. Uzdensky}
\affiliation{Center for Integrated Plasma Studies, Physics Department, 390 UCB, University of Colorado, Boulder, CO 80309, USA}

\author{Gregory R. Werner}
\affiliation{Center for Integrated Plasma Studies, Physics Department, 390 UCB, University of Colorado, Boulder, CO 80309, USA}

\author{Mitchell C. Begelman}
\affiliation{JILA, University of Colorado and National Institute of Standards and Technology, 440 UCB, Boulder, CO 80309, USA}
\affiliation{Department of Astrophysical and Planetary Sciences, 391 UCB, Boulder, CO 80309, USA}

%\date{\today}

\begin{abstract}
Nonthermal relativistic plasmas are ubiquitous in astrophysical systems like pulsar wind nebulae and active galactic nuclei, as inferred from their emission spectra.
The underlying \ac{NTPA} processes have traditionally been modeled with \pgls{FP} diffusion-advection equation in momentum space.
In this paper, we directly test the \ac{FP} framework in ab-initio kinetic simulations of driven magnetized turbulence in relativistic pair plasma.
By statistically analyzing
the motion of tracked particles,
we demonstrate the diffusive nature of \ac{NTPA} and measure the \ac{FP} energy diffusion ($D$) and advection ($A$) coefficients as functions of particle energy~$\relEnergy{}$.
We find that $D(\gamma)$ scales as $\gamma^2$
in the high-energy nonthermal tail, in line with 2nd-order Fermi acceleration theory, but
has a much weaker scaling
at lower energies.
We also find that $A$ is not negligible and reduces \ac{NTPA} by tending to pull particles towards the peak of the particle energy distribution.
This study provides strong support for the \ac{FP} picture of turbulent \ac{NTPA}, thereby enhancing our understanding of space and astrophysical plasmas.
\end{abstract}

%\maketitle

\newcommand{\citeMhdTestPtcls}{\cite{Dmitruk2003, Dmitruk2004, Kowal2012, Lynn2014, Kimura2016, Isliker2017a}}
\newcommand{\citeReconNTPA}{\cite{Zenitani2001, Jaroschek2004, Lyubarsky2008, Sironi2014, Guo2014, Guo2016, Werner2016, Werner2018, Werner2017}}
\newcommand{\citeTurbNTPA}{\cite{Kunz2016, Makwana2017, Zhdankin2017PRL, Zhdankin2018ApJL, Zhdankin2019, Comisso2018, Arzamasskiy2019}}
\newcommand{\citeShockNTPA}{\cite{Hoshino1992, Amato2006, Spitkovsky2008, Sironi2011, Marcowith2016}}
\newcommand{\citeConstrFieldsNTPA}{\cite{Arzner2006, OSullivan2009}}

\newcommand{\citeFpTheories}{\cite{Fermi1949, Kulsrud1971, Melrose1974, Skilling1975, Blandford1987, Schlickeiser1989, Miller1990, Chandran2000, Cho2006}}
\newcommand{\citeDpropSq}{\cite{Kulsrud1971, Skilling1975, Blandford1987, Schlickeiser1989, Chandran2000, Cho2006}}

\glsresetall
\section{Introduction}
Relativistic plasmas with nonthermal power-law energy distributions are ubiquitous in astrophysical systems such as
\ac{PWN} \cite{Meyer2010, Buehler2014},
jets from \ac{AGN} \cite{Begelman1984, Hartman1992}
and their radio lobes \cite{Hardcastle2009},
and
black-hole accretion-disk coronae \cite{Yuan2003}.
The underlying
\ac{NTPA} processes
have been studied theoretically for decades;
proposed mechanisms include
collisionless
shocks \cite{Blandford1987},
turbulence \cite{Kulsrud1971}, and
magnetic reconnection \cite{Hoshino2012}.
The most common turbulent \ac{NTPA} models
posit that particles gain energy in a stochastic process
(e.g., scattering off magnetic fluctuations)
that can be modeled using \pgls{FP} advection-diffusion
equation in momentum space \citeFpTheories{}.

Numerical tests of the \ac{FP} framework for \ac{NTPA} were originally performed by injecting test particles into \acl{MHD} simulations \citeMhdTestPtcls{} or artificially prescribed fields \citeConstrFieldsNTPA{}. These test-particle simulations are 
relatively inexpensive, but have
physical limitations such as
ad-hoc particle injection and the absence of
particle feedback on the fields,
which can only be resolved by considering more physically complete simulations.

Recently, first-principles kinetic (and hybrid) \ac{PIC} simulations have confirmed that
turbulence \citeTurbNTPA{}, shocks \citeShockNTPA{}, and relativistic reconnection \citeReconNTPA{} can generate efficient \ac{NTPA} in collisionless plasma.
%Since
\ac{PIC} simulations contain complete microphysical information
including the self-consistent trajectories and energy histories of individual particles.
However, this wealth of data has not yet been employed directly to
test stochastic acceleration models (e.g., \ac{FP})
or to
measure the energy diffusion and advection coefficients.

In this paper, we use tracked particles to demonstrate stochastic acceleration and directly measure the \ac{FP} coefficients in \ac{3D} \ac{PIC} simulations of driven turbulence in collisionless relativistic plasma.
Stochastic particle acceleration in relativistic plasma turbulence has important applications to astrophysical systems such as
\ac{PWN} \cite{Bucciantini2011, Tanaka2017},
\ac{AGN} accretion flows \cite{Dermer1996, Kimura2015},
\ac{AGN} jets \cite{Rieger2007, Asano2014},
and
\acl{GRB} \cite{Dermer2001}.
We consider pair plasma both for theoretical and computational simplicity, and for its relevance to high-energy astrophysical systems like \ac{PWN} and \ac{AGN} jets.
However, our methods also apply to
future investigations of \ac{NTPA} in
turbulent non-relativistic and electron-ion plasmas, as well as to other processes, e.g., magnetic reconnection.

\section{Method}
We analyze 3D simulations [performed with our \ac{PIC} code {\sc Zeltron} \cite{Cerutti2013}] of externally driven turbulence in relativistic pair plasma
\cite{Zhdankin2018ApJL}.
We focus on the largest simulation with
$1563^3$ grid cells
and 64 particles per cell (electrons and positrons combined),
totalling ${\sim}\num{\nSimPtcls}$ particles;
smaller simulations give similar results.
The simulation domain is a periodic cube of size~$L$, with
an initially uniform 
magnetic guide field $B_0 \unitvec{z}$.
The plasma is initially uniform and isotropic, with
total charged particle density $n_0$,
%combined pair density~$n_0$,
and a \MaxwellJuttner{} thermal distribution with 
a relativistically hot temperature of $T_0 \nbaeq 100 \restEnergy$,
corresponding to the average Lorentz factor
$\mean_\init \nbaapprox 3 T_0 / \restEnergy \nbaeq 300$.
The initial magnetization is $\sigma_0 = B_0^2/16\pi n_0 T_0 \nbaeq 3/8$.
In the fiducial simulation, the
normalized system size is $L/2\pi\rho_{e_0} \nbaeq 163$, where $\rho_{e_0} \nbaequiv \mean_\init \restEnergy / e B_0$ is the initial characteristic Larmor radius.
Turbulence is electromagnetically driven \cite{TenBarge2014} and becomes fully developed after several light crossing times \cite{Zhdankin2018MNRAS}, with rms turbulent magnetic fluctuations
$\delta B_\rms \nbasim B_0$.
The turbulence is essentially \Alfven{ic} \cite{Zhdankin2018MNRAS}, with initial \Alfven{} velocity
$v_{A0}/c \equiv [\sigma_0/(\sigma_0 + 1)]^{1/2} \simeq 0.52$.

Our previous studies \cite{Zhdankin2017PRL, Zhdankin2018ApJL} have shown that such turbulence reliably produces nonthermal power-law particle spectra.
In
%the remainder of
this section, we describe our procedure to investigate \ac{NTPA} in relation to the \ac{FP} framework.
First, we examine gyro-scale oscillations in particle energy and explain
their physical origin.
Then, we present our methodology
for averaging out these oscillations, which is critical for accurately measuring energy diffusion.
Finally, we detail
our tests of diffusive \ac{NTPA}
in our turbulence simulations,
and our procedure for measuring the energy diffusion and advection coefficients as
functions of particle energy.

\begin{figure}[h]
\centering
\begin{overpic}[width=0.42\textwidth]
	{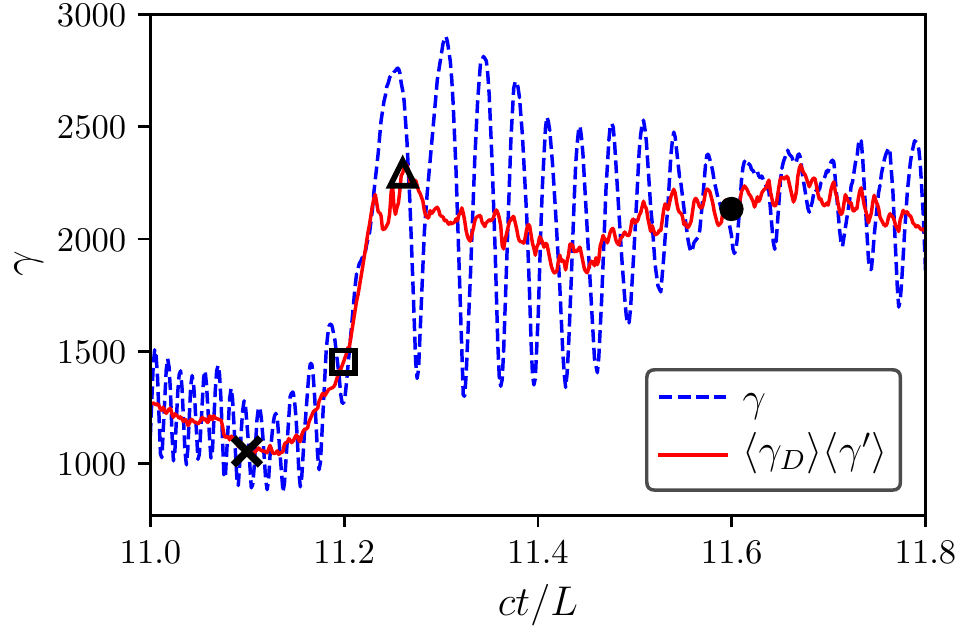}
	\put (-6, 50) {(a)}
\end{overpic}
\\ \vspace*{0.35cm}
\begin{overpic}[width=0.42\textwidth]
	{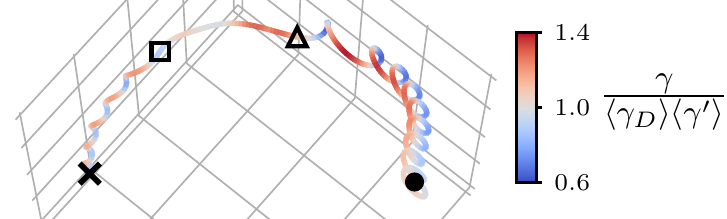}
	\put (-6, 25) {(b)}
\end{overpic}
\caption
{
(a)~Partial energy history of a tracked particle (blue dashed line), showing oscillations which are removed by our transformation (red solid line).
(b)~Trajectory of the same particle,
colored by the instantaneous ratio of the lab-frame energy to the smoothed energy.
Markers show time instances in~(a) corresponding to particle positions in~(b).
}
\label{fig:oscillation}
\end{figure}

Our analysis tracks the positions, momenta, and local electromagnetic field vectors
for a statistical ensemble of $\num{\nTrackPtcls}$ randomly-chosen particles.
We observe
order-unity
oscillations in particle energy, $\relEnergy{}$, at the gyrofrequency,
as shown for a representative particle in \figref{fig:oscillation}a.
The energy oscillates once per \gyro{orbit} (\figref{fig:oscillation}b) because of the large-scale electric field accelerating and decelerating the particle as it gyrates.
To describe this analytically,
we consider
a charged particle moving
in constant, uniform electromagnetic fields.
We use primed variables for the frame moving with the $\ExB$ drift velocity,
$\vPref$, given by $c\vPref/(c^2 + \vPrefMag^2) = \ExB/(\EsqBsq)$.

In the primed frame, where $\vec{B}'$ and $\vec{E}'$ are parallel, the particle gyrates
about $\vec{B}'$ while being accelerated along $\vec{B}'$ by~$\vec{E}'$.
Typically, $E'\ll\nobreak B'$, and so $\relEnergy{'}$
is slowly-varying on the oscillation timescale.
Then, the motion in the primed frame is approximately a simple gyration with 
$\vec{E}' \approx 0$,
and, applying the inverse Lorentz transformation, the lab-frame energy can be found:
\begin{equation}
\gamma(t) = \gamma_\pref \gamma' \left(1 +
\beta_\pref \frac{v_\perp'}{c}
\cos{\omega' t'}\right)
\label{eqn:labGamma}
,
\end{equation}
with $t = t_0 + \gamma_\pref(t' + \beta_\pref v_\perp' \sin{\omega't'}/ \omega' c)$.
Here,
$t$ is the coordinate time,
$v_\perp'$ is the particle's primed-frame velocity perpendicular to $\vec{B}'$,
$\beta_\pref \nbaeq \vPrefMag / c$,
$\gamma_\pref = \invSqrt{1 -\nobreak \vPrefMag^2/c^2}$,
$\omega' \nbaeq eB'/\gamma' \me c$ is the cyclotron frequency
with $B' \nbaeq B / \gamma_\pref$,
and
$t_0$ is a phase.
Since we are considering relativistic particles ($v_\perp' \nbasim c$) and relativistic turbulence ($E_\rms \nbasim B_0$ and $\beta_\pref \nbasim 1$),
the predicted
oscillation magnitude is comparable to~$\gamma$ (see \figref{fig:oscillation}a), and cannot be ignored.

Without further processing,
these energy oscillations are
incompatible with
a random-walk-based energy-diffusion model.
However, we aim to measure the statistical properties of \ac{NTPA}
on the \Alfven{ic} timescale ${\sim}L/v_A$ relevant to the formation of the nonthermal power law.
Since this is generally much longer than the gyroperiod,
we analyze just the secular component of the energy histories.

\newcommand{\larmorAvg}[1]{\langle #1 \rangle}
\newcommand{\vPrefAvg}{\larmorAvg{\vPref}}
Our oscillation removal procedure 
is informed by \eqnref{eqn:labGamma}, which indicates
that in the idealized case of uniform constant fields,
the secular component of the lab-frame energy is $\gamma_\pref \gamma'$.
However,
$\vPref$ fluctuates as the particle traverses small-scale fields,
so we average $\vPref$
over the
smoothed gyroperiod $\tLarDef{\gamma'}{B}$, where $\gamma' = \gamma_D \gamma (1 - \vPref \cdot \vec{v} / c^2)$ is obtained from boosting the lab-frame four-velocity $\gamma \vec{v}$ by $\vPref$.
We denote by $\vPrefAvg$ the average of $\vPref$ over this smoothed period.
We then
define the smoothed particle energy to be
$\larmorAvg{\gamma_\pref} \larmorAvg{\gamma'} $,
where
$\larmorAvg{\gamma'} = \larmorAvg{\gamma_D} \gamma (1 - \vPrefAvg \cdot\nobreak \vec{v} / c^2)$
and
$\larmorAvg{\gamma_\pref} \equiv (1 - \vPrefAvg^2/c^2)^{-1/2}$.
This transformed energy (\figref{fig:oscillation}a)
has greatly reduced oscillations.
Thus, this procedure
extracts the secular component of particle energy,
allowing us to test the \ac{FP} picture of \ac{NTPA}.
Hereafter,
$\gamma$ and ``energy'' refer to $\larmorAvg{\gamma_\pref} \larmorAvg{\gamma'}$,
except in the case of the overall particle energy distributions (\figref{fig:diffcoeff}a) and magnetic energy spectra (\figref{fig:diffcoeff}b).

We now describe our tests of
diffusive acceleration using
tracked particles.
We use $t_0 \nbaeq \startTL$ as a fiducial initial time for our measurements, by which point
a power-law particle energy spectrum has fully formed.
We then bin tracked particles by their energy at~$t_0$
in logarithmic intervals spaced by 10\%.
For each bin, we measure the
standard deviation, $\stdev$,
and the mean, $\mean$,
of the particle energy distribution as a function of subsequent times $\changeT \nbaequiv t - t_0$.
For a classical diffusion process, one expects $\stdevProptoSqrt$ if the diffusion coefficient $\Dcoeff(\gamma, t)$ varies slowly compared with $\changeT$ and $\stdev$.

We then
measure the diffusion and advection coefficients, $\Dcoeff(\gamma)$ and $\Acoeff(\gamma)$, respectively, 
for the simplest \gls{FP} equation for the energy distribution~$\fxt$, ignoring pitch angle (with respect to $\vec{B}$):
\begin{equation}
\pdt f = \pdx (\Dcoeff \pdx f) - \pdx (\Acoeff f)
.
\label{eqn:fp}
\end{equation}

Limiting our 
measurements to times where
$\Delta t \lesssim\nobreak{}L/v_A$,
$\stdev \ll \bcGamma$,
and
$\Delta\mean \nbaequiv \mean(t) -\nobreak\mean(t_0) \ll \bcGamma$,
we approximate the bin distribution as narrow and the coefficients as constant in time.
Applying \eqnref{eqn:fp}, we find:
\begin{align}
\Delta\mean(\bcGamma, \changeT) &= [\pdx \Dcoeff |_\bcGamma + A(\bcGamma)] \changeT \equiv \Mcoeff(\bcGamma) \changeT
\label{eqn:meant} \\
\stdev(\bcGamma, \changeT) &= \sqrt{2 \Dcoeff[(\bcGamma)] \changeT}
.
\label{eqn:stdevt}
\end{align}
We first measure $\Dcoeff(\gamma)$ and $\Mcoeff(\gamma)$
by applying \eqnsref{eqn:meant}{eqn:stdevt} to each 
energy bin and then calculate
$\Acoeff(\gamma) \nbaequiv \Mcoeff - \Dcontrib$.

\section{Results}
We first describe the evolution of the overall lab-frame distribution,~$\fx$.
Starting from a thermal distribution, $\fx$ acquires a 
power-law tail extending to the
system-size limit,
$\gamma_{\rm max} \nbaequiv L e B_0 / 2 \restEnergy \nbasimeq \num{1.5e5}$, and
gradually hardening over
time\textemdash{}its index converges to approximately -3 by $12.3L/c$.
At the start 
of our measurements, $t_0=\startTL$, the 
index is approximately $-3.2$,
the peak of $\fx$ is 
at $\gammaPeak \simeq 520$,
and the mean at $\gammaAvg \simeq 1170$.
Because the system lacks an energy sink,
$\gammaPeak$ and $\gammaAvg$ increase at a rate of about $40\invLC$ and $100\invLC$, respectively.

\newcommand{\numPtcl}{{\sim}2200}
\newcommand{\bcVal}{\num{5e3}}
We now present tests of energy diffusion. 
For illustration,
\figref{fig:hist}a shows
the evolution of the energy distribution of a single bin of particles
with $\bcGamma \nbaeq \bcVal$,
deep in the power-law section.
We find that $\mean \propto \changeT$ for small $\changeT$ (\figref{fig:hist}b),
while $\stdevProptoSqrt$ (\figref{fig:hist}c), consistent with simple diffusion.

\newcommand\bsl{solid}
\newcommand\bdl{dashed}
\newcommand{\compSqrt}{(L/c\Delta t)^{1/2}}
\begin{figure}[h]
\centering
\begin{overpic}[width=0.45\textwidth]
	{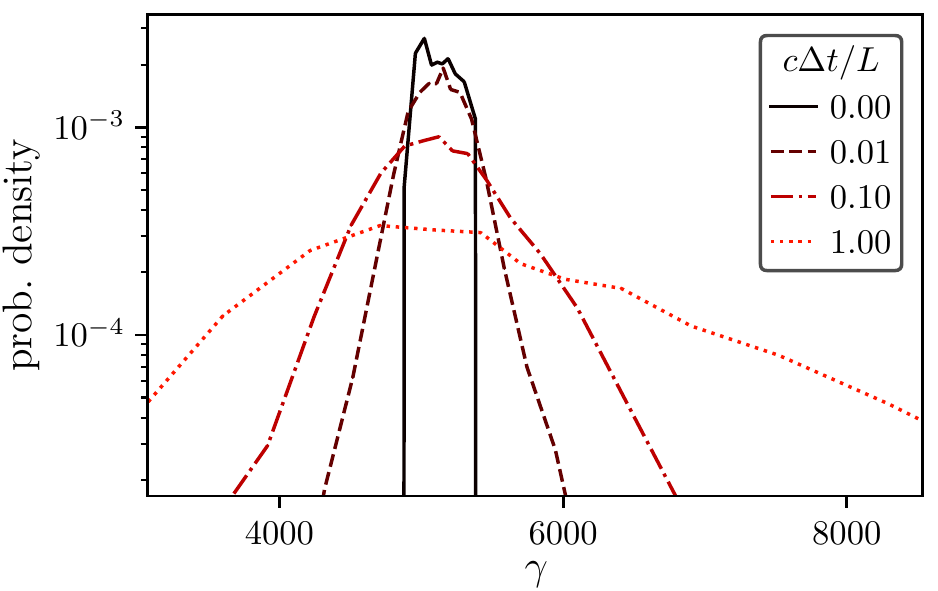}
	\put (20, 54) {(a)}
\end{overpic}
\\ \vspace*{0.2cm}
\begin{overpic}[width=0.47\textwidth]
	{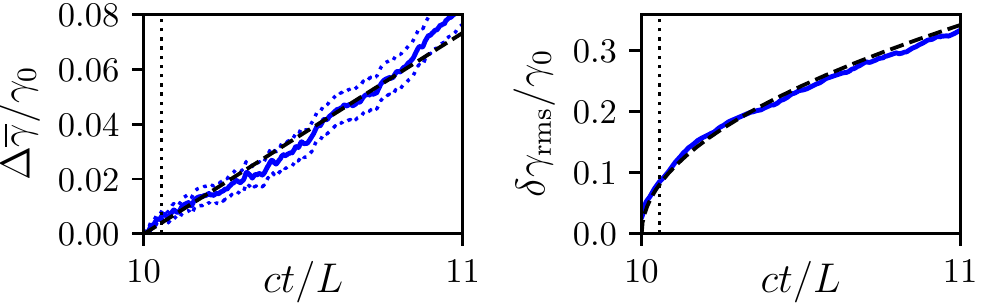}
	\put (19, 24) {(b)}
	\put (70, 24) {(c)}
\end{overpic}
\caption
{
(a) Time evolution of the energy distribution
for a bin of $N \numPtcl{}$ particles
with bin-center energy
$\bcGamma = \bcVal$.
For this bin, (b)~shows
$\mean(t)/\bcGamma$
(\bsl) with standard error ranges $(\mean(t) \pm \stdev(t) / \sqrt{N}) / \bcGamma$ (dotted) and a linear $\changeT$ fit (\bdl),
and (c)~shows
$\stdev(t)/\bcGamma$
(\bsl) with a $\sqrt{\changeT}$ fit (\bdl).
In (b) and (c), a vertical dotted line
is placed at
$\changeT \nbaeq \tLarmor(\bcGamma, B_\rms)$,
the gyroperiod corresponding to $\bcGamma$.
}
\label{fig:hist}
\end{figure}

\begin{figure}[h]
\centering
\includegraphics[width=0.48\textwidth]{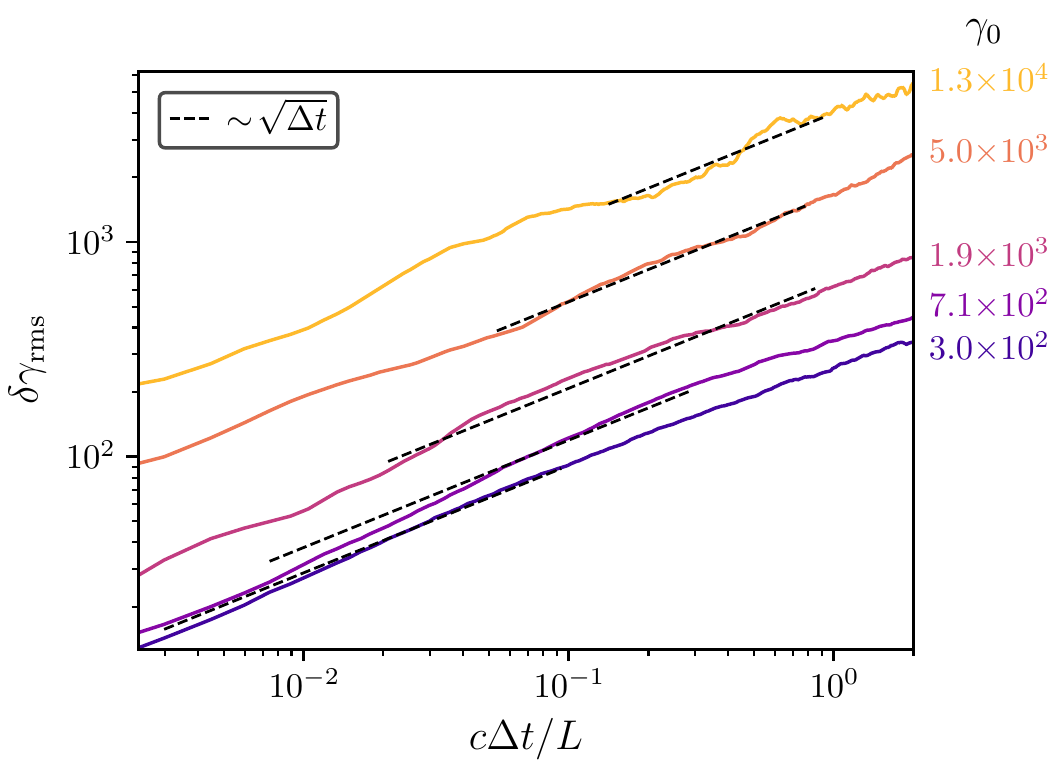}
\caption
{
The standard deviation, $\stdev$, of the particle energies in several bins (solid lines), with corresponding $\sqrtT$ fits (dashed lines), each annotated by the initial bin-center energy $\bcGamma$.
Each fit is drawn over the corresponding fitting interval, which begins at
$\Delta t \nbaeq \tLarmor(\gamma_0, B_\rms)$, the gyroperiod for the bin-center energy of the corresponding bin.
}
\label{fig:stdevt}
\end{figure}

\figref{fig:stdevt} shows $\stdev(\changeT)$ for several bins
along with corresponding $\sqrtT$ fits.
To avoid artifacts of the smoothing procedure,
each fit begins after 
one gyroperiod $\tLarmor(\gamma_0, B_\rms) \nbaequiv \tLarDef{\gamma_0}{B_\rms}$. 
To ensure \eqnsref{eqn:meant}{eqn:stdevt} are valid, each fit ends
when $\relStdev$ reaches 0.3, $\Delta \relMean$ reaches 0.1, or $\changeT \nbaeq 2L/c$,
whichever is earliest.
Under these criteria, 
almost all fits end before $\Delta t = 1L/c$.
The fits generally agree well with the data over the fitted intervals.
While some of the plotted $\stdev(\changeT)$ have intervals of weakly anomalous energy diffusion
these regimes are not our current focus,
and we refer to studies of anomalous energy diffusion in plasma turbulence \cite{Isliker2017a, Isliker2017b}.
In summary, \figsref{fig:hist}{fig:stdevt}
confirm our expectations of
a standard diffusive process,
supporting
the \ac{FP} model of turbulent \ac{NTPA}.

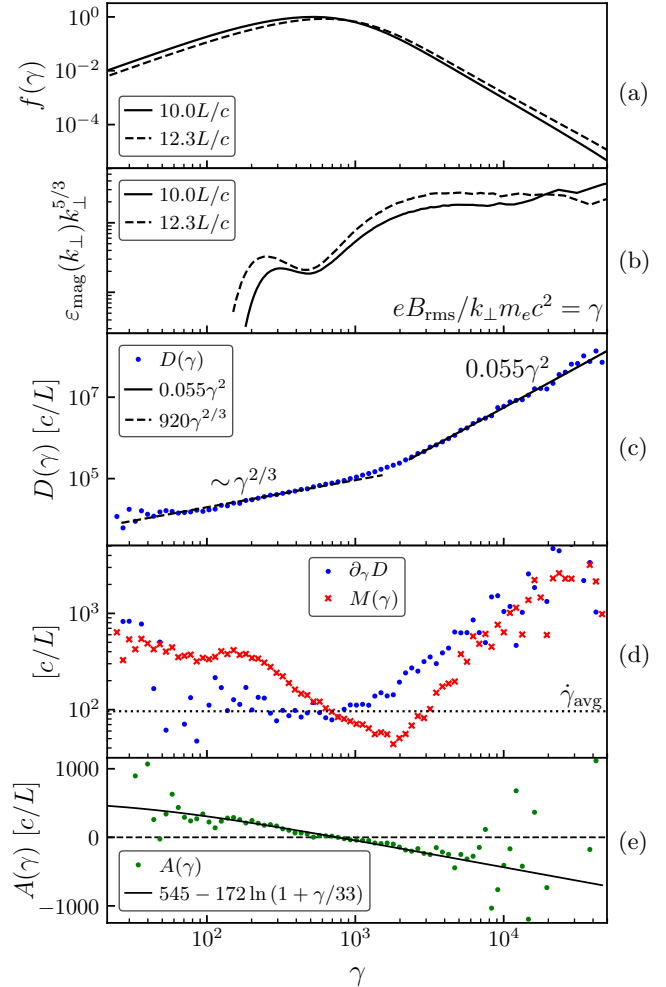
\begin{figure}[h]
\raggedright
\begin{overpic}[width=0.44\textwidth]
	{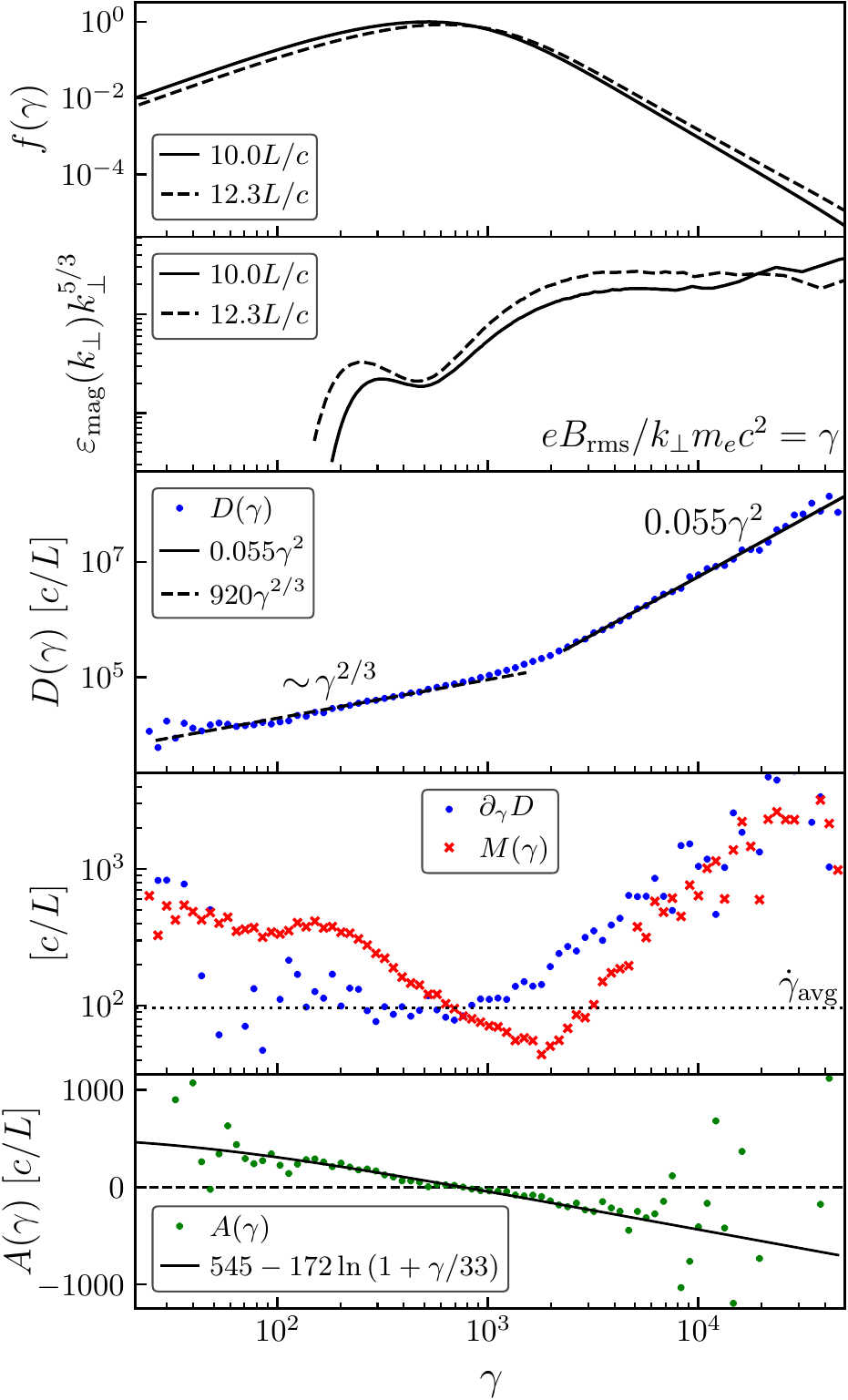}
	\put (61.5, 90) {(a)}
	\put (61.5, 73) {(b)}
	\put (61.5, 54) {(c)}
	\put (61.5, 33) {(d)}
	\put (61.5, 14) {(e)}
\end{overpic}\\
\caption
{
(a)~The overall particle energy distribution $\fx$
at the start of the measuring interval ($\teq{\startTL}$, solid line), and a short time later ($\teq{12.3L/c}$, dashed line).
(b)~The magnetic energy spectrum $\epsilon_\mathrm{mag}(k_\perp)$ compensated by $k_\perp^{5/3}$ 
vs.\ $eB_\rms / k_\perp \me c^2$, the Lorentz factor corresponding to a perpendicular gyroradius equal to
the inverse of the perpendicular (to $\vec{B_0}$) wavenumber $k_\perp$.
(c)~The diffusion coefficient $\Dcoeff(\gamma)$ (dots),
with power-law fits of index 2 (solid line) in the nonthermal region, and
index \lowPower{} (dashed line) in the low-energy region.
(d)~The acceleration rate $\Mcoeff(\gamma)$ (crosses)
and the contribution to $\Mcoeff(\gamma)$ by $\Dcontrib$ (dots).
For reference, the overall average rate of energy gain 
$\dot{\gamma}_\mathrm{avg} \simeq 100 \invLC$ is shown as a dotted line.
(e)~The advection coefficient $\Aeqn$ (dots) with logarithmic fit (solid line).
}
\label{fig:diffcoeff}
\end{figure}

\newcommand{\DpropGammaSq}{\Dcoeff \propto \gamma^2}
We now report on our measurements of
$\Dcoeff(\gamma)$ and~$\Acoeff(\gamma)$.
\figref{fig:diffcoeff}c shows $\Dcoeff(\gamma)$, extracted from the fits of $\stdev(\changeT)$ using \eqnref{eqn:stdevt}.
In the high-energy nonthermal tail ($\num{2e3} \lesssim\nobreak \gamma \lesssim\nobreak \num{3e4}$, see \figref{fig:diffcoeff}a),
$\Dcoeff \nbaeq \Dmeas (\invLC) \gamma^2$
is an excellent fit,
while for lower energies $\gamma \lesssim \gammaPeak$,
there is a much shallower scaling
roughly consistent with $\Dcoeff \propto \gamma^{\lowPower}$.
We observe that $\DpropGammaSq$ for particles gyroresonant with fluctuations in the inertial range of the magnetic energy spectrum (\figref{fig:diffcoeff}b),
while the lower-energy scaling corresponds to the sub-inertial range of turbulence.
In simulations with different magnetization, system size, and number of particles per cell (not shown), the high-energy scaling of $\DpropGammaSq$ is maintained while the low-energy behavior varies slightly.

The high-energy scaling of $\DpropGammaSq$
is commonly predicted by \ac{NTPA} theories \citeDpropSq.
We compare our high-energy fit,
$\Do \nbaeq \Dmeas \invLC$,
to the theoretical prediction
from 2nd-order Fermi acceleration,
$\Do \nbaeq u_A^2/3 c \mfp$,
for ultra-relativistic particles interacting with isotropic scatterers moving at the \Alfven{} velocity, where
$u_A = v_A / (1 - v_A^2 / c^2)^{1/2}$ and
$\mfp$ is the mean free path between scattering events~\cite{Blandford1987, Longair2011}.
At $t_0 \nbaeq \startTL$, $v_A \nbaeq 0.51c$, giving a theoretical scaling of $\Do \nbaeq 0.12 c/\mfp$,
which agrees with our fit if $\mfp \nbasim 2L$.
We also find, from simulations with reduced driving wavelength (not shown), that the turbulence driving scale, and not the box size, sets $\Do$ (and hence $\mfp$).
While the value of $\Do$ is consistent with 
2nd-order Fermi acceleration from gyroresonant scattering of particles by \Alfven{} modes,
further work is needed to test this view. This includes studying the effect of varying magnetization, as well as the spatial transport characteristics.

To compare the effects of the
first- and second-order derivative
terms in \eqnref{eqn:fp}, we separate the contributions of $\Acoeff$ and $\Dcontrib$ to the average acceleration rate $\Mcoeff(\gamma) = \pdt\mean$.
We extract $\Mcoeff(\gamma)$ from linear fits of $\mean(\changeT)$ [see \eqnref{eqn:meant}],
using the same time intervals as those used for fitting $\stdev(\changeT)$ to measure~$\Dcoeff(\gamma)$.
As shown in \figref{fig:diffcoeff}d, $\Mcoeff(\gamma)$ is positive, as expected with external energy injection, and has a minimum near $\gammaAvg \nbaapprox 1200$.

We then compute the contribution of energy diffusion to acceleration, $\Dcontrib$, (\figref{fig:diffcoeff}d) and subsequently the advection coefficient $\Aeqn$ (\figref{fig:diffcoeff}e).
In the high-energy power-law section ($\gamma > \gammaPeak$), $\Acoeff$ is negative, while for low energies ($\gamma < \gammaPeak$), $\Acoeff$ is positive.
Overall, $\Acoeff$ tends to pull particle energies towards $\gammaPeak$, narrowing~$\fx$.
We find that $\Acoeff$ is reasonably approximated by a logarithmic scaling with energy.
For a momentum-space \ac{FP} equation containing only diffusion [see, e.g., \citet{Ramaty1979}], the expected energy-space advection coefficient for \eqnref{eqn:fp} is $A = 2D/\gamma$. The measured negative logarithmic scaling thus implies that the momentum-space \ac{FP} equation contains a significant advection term.
In the nonthermal section, the magnitude of $A$ is generally smaller than, but still comparable to that of $\Dcontrib$.
Hence, the evolution of the nonthermal population in this simulation cannot be interpreted as being due to $D$ alone.
This also complicates estimates of the acceleration time based on $\Do$.
The rate of overall energy increase from advection is $\int d\gamma Af \approx -23c/L$ while that from diffusion is $\int d\gamma (\Dcontrib) f \approx 140c/L$.
The general effects of systematic and stochastic acceleration (due to $\Acoeff$ and $\Dcoeff$ respectively) are discussed in, e.g., \citet{Pisokas2018} and \citet{Vlahos2019}.
We note that our measurement of $\Acoeff$ has considerable scatter outside of the central range, $10^2 \lesssim\nobreak \gamma \lesssim\nobreak 10^4$, as it depends on the difference between two noisy quantities.

\newcommand{\tA}{4.5}
\newcommand{\tB}{8.9}
\newcommand{\tC}{13.4}

\newcommand{\ctL}{ct/L}
\begin{figure}[h]
\centering
\begin{overpic}[width=0.44\textwidth]
	{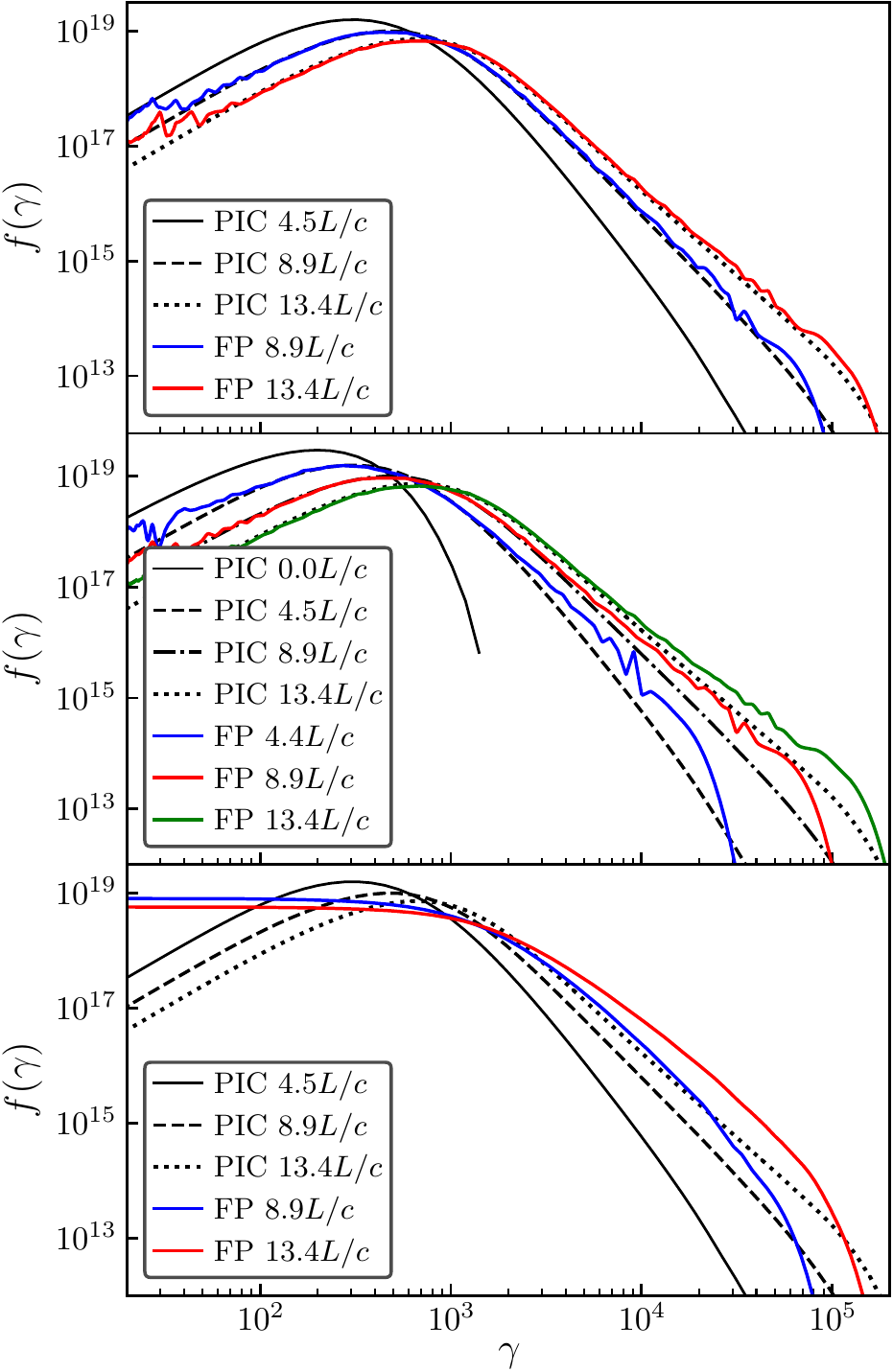}
	\put (59, 93) {(a)}
	\put (59, 62) {(b)}
	\put (59, 31) {(c)}
\end{overpic}\\
\caption
{
(a) Particle distributions $\fx$ from the \ac{PIC} simulation
at $\ctL \in \{\tA, \tB, \tC\}$
(black solid, dashed, and dotted lines, respectively), 
and the \ac{FP} solution (initialized at
$\ctL \nbaeq \tA$)
at $\ctL \in \{\tB, \tC\}$
(red and blue solid lines, respectively).
(b) Similar to (a) but for an \ac{FP} solution initialized at $\ctL \nbaeq 0$.
(c) Similar to (a) but with the advection coefficient artificially set to zero.
}
\label{fig:fpevolution}
\end{figure}

Finally, we test whether the \ac{FP} equation with coefficients measured by our methodology can reproduce the evolution of $\fxt$ from the \ac{PIC} simulations.
We first repeat the measurements of $\Dcoeff$ and $\Acoeff$ at almost 200 different $t_0$ evenly spaced over the entire simulation, thus obtaining time-dependent coefficients.
We then insert the measured coefficients into the \ac{FP} equation [\eqnref{eqn:fp}] and solve it numerically using a finite-volume method.
We use linear interpolation in $\gamma$ and nearest-neighbor interpolation in $t$.
The \ac{FP} coefficients are extrapolated as constant in $\gamma$ for energies without enough particles to measure them.
Comparing with the particle energy distributions obtained from the \ac{PIC} simulation, we find that the \ac{FP} equation gives excellent agreement at all subsequent times when initialized with the corresponding \ac{PIC} distribution as early as $\teq \tA L/c$ (\figref{fig:fpevolution}a), with moderate errors if initialized at $\teq 0$ (\figref{fig:fpevolution}b).
This may be due to the turbulence not being fully developed at early times.
Finally, evolving from $\teq \tA L/c$ with $\Acoeff$ set to zero shows a clear mismatch with the \ac{PIC} distributions (\figref{fig:fpevolution}c), giving us confidence that the measurements of $\Acoeff$ are reasonably accurate and that $\Acoeff$ significantly affects the evolution of $\fxt$.
Thus, in this regime of fully-developed turbulence, we find that the \ac{FP} equation with our coefficient measurement methodology is appropriate for modeling \ac{NTPA}. 
We leave further investigation of the early-time behavior to future work.

\glsresetall
\section{Conclusions}
In this study we rigorously demonstrate, for the first time,
diffusive \ac{NTPA} in first-principles \ac{PIC} simulations of driven relativistic plasma turbulence,
through direct statistical measurements using large numbers of tracked particles.
We introduce a procedure to suppress large-amplitude \gyro{oscillations} of particle energy, which is critical for revealing the diffusive nature of \ac{NTPA} and measuring the \ac{FP} coefficients.
We find that the energy diffusion coefficient $\Dcoeff$ scales
with particle energy $\relEnergy{}$
as $\Dcoeff \nbasimeq 0.06(c/L)\gamma^2$ in the high-energy nonthermal power-law region,
in line with theoretical expectations \citeDpropSq,
while there is a much shallower scaling
at energies below the peak of the energy distribution.
We also measure
the energy advection coefficient $\Acoeff(\gamma)$, though with more uncertainty. We find that $\Acoeff$ is not negligible, and tends to narrow the distribution by accelerating low-energy particles and decelerating high-energy particles.
Furthermore, a numerical solution of
the \ac{FP} equation with the measured coefficients
reproduces the evolution of the particle energy spectrum from the \ac{PIC} simulation
over a significantly longer time interval than was
used for measuring the coefficients.
This suggests that this simple \ac{FP}
model can fully account for \ac{NTPA} in our simulations.
These results thus lend strong first-principles numerical support to a broad class of turbulent \ac{NTPA} theories.

Our new methodology
can also be applied to future tracked-particle studies of \ac{NTPA}
in other contexts such as shocks or magnetic reconnection,
and over broader ranges of physical regimes.
Future work may investigate the effects of
system parameters such as magnetization, plasma beta, and guide field strength;
radiative cooling;
relativistic vs non-relativistic regimes; and
plasma composition (e.g., pair vs electron-ion plasma).
In addition,
the analysis can be extended to include pitch-angle dependence and scattering.
Characterizing stochastic \ac{NTPA} in various regimes has important implications in a broad range of contexts such as
solar flares, the solar wind, \acl{PWN}, \acl{AGN}, \acl{GRB}, and cosmic ray acceleration in supernova remnants.
Thus, this study will facilitate
more detailed tests of \ac{NTPA} theories against \ac{PIC} simulations exploring various physical situations,
thereby advancing our understanding of space, solar, and high-energy astrophysical phenomena.

\acknowledgments
The authors acknowledge support from NSF grants AST-1411879 and AST-1806084, and NASA ATP grants NNX16AB28G and NNX17AK57G.
An award of computer time was provided by the Innovative and Novel Computational Impact on Theory and Experiment (INCITE) program. This research used resources of the Argonne Leadership Computing Facility, which is a DOE Office of Science User Facility supported under Contract DE-AC02-06CH11357.

\newcommand\rpp{Rep. Prog. Phys.}

\providecommand{\noopsort}[1]{}

\end{document}